\documentclass[aps,prb,showpacs,twocolumn,floats,epsfig]{revtex4}
\usepackage{amssymb}
\usepackage{amsbsy}
\usepackage{amsmath}
\usepackage{epsfig}
\newcommand\beq{\begin{equation}}
\newcommand\eeq{\end{equation}}
\newcommand\bea{\begin{eqnarray}}
\newcommand\eea{\end{eqnarray}}
\newcommand\al{\alpha}
\newcommand\ga{\gamma}
\newcommand\de{\delta}

\newcommand\si{\sigma}
\newcommand\dg{\dagger}
\newcommand\pa{\partial}

\newcommand\non{\nonumber}
\newcommand\bib{\bibitem}

\begin{document}

\title{Junction between surfaces of two topological insulators}

\author{Diptiman Sen and Oindrila Deb}

\affiliation{Center for High Energy Physics, Indian Institute of Science, 
Bangalore 560 012, India}


\begin{abstract}
We study the properties of a line junction which separates the surfaces of 
two three-dimensional topological insulators. The velocities of the Dirac 
electrons on the two surfaces may be unequal and may even have opposite signs.
For a time reversal invariant system, we show that the line junction is 
characterized by an arbitrary 
parameter $\al$ which determines the scattering from the junction. If the 
surface velocities have the same sign, we show that there can be edge states 
which propagate along the line junction with a velocity and orientation of 
the spin which depend on $\al$ and the ratio of the velocities. Next, we 
study what happens if the two surfaces are at an angle $\phi$ with respect 
to each other. We study the scattering and differential conductance through 
the line junction as functions of $\phi$ and $\al$. We also find that there 
are edge states which propagate along the line junction with a velocity and 
spin orientation which depend on $\phi$. Finally, if the surface velocities 
have opposite signs, we find that the electrons must transmit into the 
two-dimensional interface separating the two topological insulators. 
\end{abstract}


\maketitle

\section{Introduction}
\label{sec1}

Recent years have witnessed extensive theoretical 
\cite{kane1,bern,kane2,moore,qi1,roy,egger} and experimental 
\cite{konig1,konig2,hsieh1,xia,chen,hsieh2,rous,hsieh3,zhang2,zhang1}
studies of a class of two-and three-dimensional materials called topological 
insulators; for reviews, see Refs. \onlinecite{hasan} and \onlinecite{qi2}.
A topological insulator (TI) is a material which is gapped in the bulk but has 
gapless states at the surface which is one- or two-dimensional if the TI is 
two- or three-dimensional. Further, the electrons have strong spin-orbit 
coupling, and the surface states are described by a massless Dirac equation 
(in one or two dimensions) in which the directions of the spin angular 
momentum and linear momentum are tied to each other; they both lie in the 
plane of the surface and they are perpendicular to each other. The 
three-dimensional TIs
come in two classes, strong and weak, which respectively have an odd and even 
number of massless Dirac cones at the surface \cite{bern,kane2,moore,roy}. 
In strong TIs, the gaplessness of an odd number of Dirac cones is protected 
by time reversal invariance; time reversal invariant perturbations such as 
non-magnetic disorder and lattice imperfections do not produce a gap, while 
time reversal breaking terms such as magnetic impurities or an external 
magnetic field can produce a gap for the surface states. For materials such 
as $\rm HgTe$, ${\rm Bi_2 Se_3}$ and ${\rm Bi_2 Te_3}$, specific surfaces 
have been found which have a single Dirac cone at one point of the 
two-dimensional Brillouin zone \cite{konig1,hsieh1,xia,hsieh2,rous,hsieh3}. 
Several features of these surface Dirac electrons have been investigated
recently. These include the existence of Majorana fermion modes at a 
magnet-superconductor interface \cite{kane4,been,tanaka1,linder}, anomalous 
magnetoresistance of ferromagnet-ferromagnet junctions \cite{tanaka2}, 
realization of a switch by magnetically tuning the transport of the electrons 
by a proximate ferromagnetic film \cite{mondal}, spin textures with chiral 
properties \cite{hsieh2,rous,hsieh3}, and the anisotropy of spin polarized 
scanning tunneling microscope (STM) tunneling into the surface of a TI 
\cite{saha}. Some general 
properties of the surface states and their scattering at the edges of two- 
and three-dimensional TIs have been studied in Ref. \onlinecite{jiang}.

A recent paper \cite{taka} has studied what happens when two three-dimensional
TIs are brought close to each other so that their surfaces touch each other
along a line; we will call this a line junction. This produces
a system in which two two-dimensional surfaces governed by massless Dirac 
equations, with possibly different velocities $v_1$ and $v_2$, share a 
one-dimensional boundary; similar studies have been carried out in Refs. 
\onlinecite{raoux} and \onlinecite{concha} and in one dimension in Ref. 
\onlinecite{peres}. If a plane wave is incident on the boundary from one
side with an angle of incidence $\theta$, it will be reflected and transmitted 
with certain amplitudes which depend on $\theta$ and the ratio of velocities 
$v_1/v_2$. Ref. \onlinecite{taka} made the interesting observation that if 
the velocities $v_1$ and $v_2$ have the same sign, the scattering problem can 
be solved in a straightforward way, but if they have opposite velocities, 
there is a difficulty; at least for the case of normal incidence, conservation
of spin implies that the reflection and transmission amplitudes must both 
vanish. The way out of this difficulty is to consider two more sets of 
surface states which lie at the interface of the two TIs; then these states 
necessarily have some gapless states, even if one takes into account tunneling
between them. Ref. \onlinecite{taka} proposed that a low-energy plane wave 
which is incident normally on the line junction must get transmitted into 
these interface states.

In this paper, we will generalize the study of a junction between two TI 
surfaces in several ways. In Sec. II, we will consider the case of velocities 
with the same sign. We find that the most general condition at the line 
junction which conserves the current has one arbitrary parameter $\al$ if
we assume time reversal invariance. (The
condition assumed in Ref. \onlinecite{taka} corresponds to the special case 
of $\al = 0$). We will compute the reflection and transmission amplitudes 
across the line junction as functions of $\theta$, $\al$ and the ratio of the 
velocities on the two sides of the line junction. We will also show that 
depending on the values of $\al$ and the ratio of the two velocities, there 
can be edge states which travel along the line junction with a momentum $k$
but decay exponentially perpendicular to it. The velocity of the edge
states along the junction is smaller than the velocities of the plane
waves propagating far from the junction, while the spin orientation of 
the edge states has a component pointing out of the surface in contrast to
the plane waves whose spins lie in the surface. In Sec. III, we will
consider a TI which has two surfaces which 
meet at a line junction at an angle $\phi$; we will show that this 
provides an interesting scattering problem in which the reflection and 
transmission amplitudes depend on $\theta$, the bending angle $\phi$ and
the parameter $\al$. We will use the transmission probability to compute the 
differential conductance across the line junction. Further, we will show that 
for $\phi \ne 0$, there are edge states which travel along the line junction 
with a momentum $k$ but decay exponentially away from the junction; the 
velocity and the spin orientation of these states depend on $\phi$. 
In Sec. IV, we will examine the case of two TI surfaces where the 
velocities have opposite signs, and we will introduce an interface with
two additional surfaces. We will explicitly derive the 
energy spectrum and wave functions at the interface assuming a simple form
of the tunneling between the interface surfaces. We will then demonstrate that 
a wave incident on the surface indeed transmits into these states although 
the reflection and the transmission amplitudes on the surface are generally 
non-zero if the angle of incidence $\theta \ne 0$. In Sec. V, we will point 
out some ways by which the edge states moving along the junction may be 
experimentally detected, and we will make some concluding remarks. In the
Appendix, we will study the most general boundary condition which is 
consistent with the Hermiticity of the Hamiltonian and the conservation
of the current at the junction in the model discussed in Sec. II. We will 
show that the general condition involves four real parameters. However, if 
we assume that the system is invariant under time reversal, then there can 
be only one parameter; physically this corresponds to the strength of a 
potential barrier which may be present at the junction.

\section{Junction between surface velocities with same sign}
\label{sec2}

\begin{figure} \begin{center} \epsfig{figure=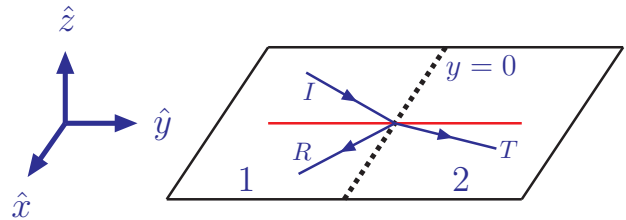,width=8.6cm} \end{center}
\caption{(Color online) Schematic picture of two regions of the $x-y$ plane, 
labeled 1 and 2, which are separated by a line junction at $y=0$ where a 
thin barrier may be present. A wave 
incident ($I$) from region 1 and the corresponding reflected ($R$) and 
transmitted ($T$) waves are shown.} \label{fig1} \end{figure}

In this section, we will study a system with a line junction which separates
the surfaces of two three-dimensional TIs, labeled 1 and 2. The surfaces
define the $x-y$ plane, and the line junction lies along $y=0$, as shown in 
Fig.~\ref{fig1}; a thin barrier may be present along $y=0$ as we will 
discuss below. The Hamiltonians in the two regions, $y< 0$ and $y>0$, are 
given by the two-dimensional massless Dirac form
\bea H_1 &=& - i v_1 ~( \si^x \pa_y ~-~ \si^y \pa_x ), \non \\
H_2 &=& -i v_2 ~( \si^x \pa_y ~-~ \si^y \pa_x ), \label{ham1} \eea
where $\si^a$ denote Pauli matrices, and $v_1$, $v_2$ denote the Fermi 
velocities on the two surfaces respectively. We will assume here that $v_1, 
v_2 > 0$. The Hamiltonians $H_i$ in Eq.~(\ref{ham1}) act on wave functions 
$\psi_i$, where $i=1$ and 2 for $y<0$ and $>0$, respectively. (We will 
generally set $\hbar =1$ everywhere). 

Hamiltonians defined in disjoint regions, such as the ones given in 
Eq.~(\ref{ham1}), 
do not define a system completely; this has been emphasized earlier for the 
Dirac equation in one dimension \cite{peres}. To define the system fully, 
we need to specify the boundary conditions that the wave functions must 
satisfy at $y=0$ in order to ensure that the Hamiltonian $H$ of the entire 
system, given by $H_1$ for $y<0$ and $H_2$ for $y >0$, is Hermitian, namely,
that
\bea && \int_{-\infty}^\infty ~dx ~\left[ \int_{-\infty}^{0-} dy ~+~ 
\int_{0+}^{\infty} dy \right] ~\chi'^\dg ~H ~\chi \non \\
&=& \int_{-\infty}^\infty ~dx ~\left[ \int_{-\infty}^{0-} dy ~+~ 
\int_{0+}^{\infty} dy \right] ~(H \chi')^\dg ~\chi \label{rel1} \eea
for any two wave functions $\chi$ and $\chi'$ which satisfy the boundary
conditions. On doing an integration by parts with $\pa_y$, we find that 
Eq.~(\ref{rel1}) requires
\beq v_1 ~((\chi'_1)^\dg \si^x \chi_1)_{y \to 0-} ~=~ v_2 ~((\chi'_2)^\dg \si^x
\chi_2)_{y \to 0+}. \label{rel2} \eeq

We may now ask: what is the most general linear relation between $(\chi_1)_{y 
\to 0-}$ ($(\chi'_1)_{y \to 0-}$) and $(\chi_2)_{y \to 0+}$ ($(\chi'_2)_{y \to
0+}$) which satisfies Eq.~(\ref{rel2})? The answer to this is discussed in 
detail in the Appendix; we show there that the general linear relation involves
four arbitrary real parameters, and we provide a physical understanding of 
these parameters. We then argue that if the system is invariant under time 
reversal, the linear relation involves only one real parameter $\al$ and is 
given by
\beq (\chi_2)_{y \to 0+} ~=~ \sqrt{\frac{v_1}{v_2}} ~e^{- i\al \si^x} ~
(\chi_1)_{y \to 0-}, \label{bc2} \eeq
and similarly between $(\chi'_2)_{y \to 0+}$ and $(\chi'_1)_{y \to 0-}$.
We note that changing $\al \to \al + \pi$ has no effect on any physical 
quantities (such as reflection and transmission probabilities) since this is 
just equivalent to changing $\psi_2 \to - \psi_2$. In the rest of this paper, 
we will assume that $\al$ lies in the range $-\pi/2 \le \al \le \pi/2$. 
In Refs. \onlinecite{taka,raoux,concha,peres}, the relation between the 
wave functions at the line junction was taken to be
\beq (\chi_2)_{y \to 0+} ~=~ \sqrt{\frac{v_1}{v_2}} ~(\chi_1)_{y \to 0-}. \eeq
Clearly this is a special case of Eq.~(\ref{bc2}) corresponding to $\al =0$.

Given the form of the Hamiltonians in Eq.~(\ref{ham1}), the current density 
${\vec J}_i = (J_{ix},J_{iy})$ on surface $i$ ($=1,2$) is given by
\bea {\vec J}_i ~=~ v_i ~(-\psi_i^\dg \si^y \psi_i, \psi_i^\dg \si^x \psi_i). 
\label{curr} \eea
[This can be derived by using the equation of motion $i \pa_t \psi_i = H_i 
\psi_i$ and the equation of continuity $\pa_t \rho_i + {\vec \nabla} \cdot 
{\vec J}_i = 0$, where the charge density is $\rho_i = \psi_i^\dg \psi_i$.]
Current conservation in the $\hat y$ direction at the line junction at $y=0$ 
implies that
\beq v_1 ~(\psi_1^\dg \si^x \psi_1)_{y \to 0-} ~=~ v_2 ~(\psi_2^\dg \si^x 
\psi_2 )_{y \to 0+}, \label{curr1} \eeq
We now see that this is ensured by the same condition (\ref{bc2}) which
ensures Hermiticity of the Hamiltonian $H$.

The system described by Eqs. (\ref{ham1}) and (\ref{bc2}) is invariant under 
two discrete symmetries, namely, parity and time reversal. Under parity, 
$x \to -x$ and $\psi \to \si^x \psi$. Under time reversal, $t \to -t$ and 
$\psi \to \si^y \psi^*$.

\subsection{Scattering}

For an electron far from the line junction, the eigenstates of the Hamiltonian
are plane waves labeled by a momentum ${\vec k} =(k_x,k_y)$. For a velocity 
$v_i > 0$, the energy spectrum and wave functions are given by
\bea \psi_+ &=& \frac{1}{\sqrt 2} ~\left( \begin{array}{c}
1 \\
e^{-i\theta} \end{array} \right) ~e^{i(k_x x + k_y y - E_+ t)}, \non \\
\psi_- &=& \frac{1}{\sqrt 2} ~\left( \begin{array}{c}
1 \\
-e^{-i\theta} \end{array} \right) ~e^{i(k_x x + k_y y - E_- t)}, 
\label{psien1} \eea
for $E_+ = v_i k$ and $E_- = - v_i k$ respectively, where $\cos \theta = k_y /
k$, $\sin \theta = k_x /k$, and $k=\sqrt{k_x^2 + k_y^2}$. Note that the 
directions of the spin and momentum are tied to each other; for positive energy
states, they are related by ${\vec \si} = {\vec k} \times {\hat z}$, where 
$\hat x$, $\hat y$ and $\hat z$ form a right-handed coordinate frame. 

Now let a plane wave be incident on the line junction from the left ($y < 0$)
with momentum ${\vec k}_I = (k_{Ix},k_{Iy})$ and positive energy $E_+ = 
v_1 |{\vec k}_I|$. The reflected and transmitted waves will have momenta 
${\vec k}_R = (k_{Rx},k_{Ry})$ and ${\vec k}_T = (k_{Tx},k_{Ty})$ and
amplitudes $r$ and $t$ respectively. We can determine all these quantities
in terms of the incident momentum ${\vec k}_I$ using conservation of energy, 
momentum in the $\hat x$ direction (due to translation invariance in that 
direction), and current conservation in the $\hat y$ direction at the line 
junction. Conservation of energy implies that $v_1 \sqrt{k_{Ix}^2 + k_{Iy}^2} 
= v_1 \sqrt{k_{Rx}^2 + k_{Ry}^2} = v_2 \sqrt{k_{Tx}^2 + k_{Ty}^2}$, while 
conservation of momentum in the $\hat x$ direction implies that 
\beq k_{Ix} ~=~ k_{Rx} ~=~ k_{Tx}. \eeq
These equations imply that
\bea k_{Ry} &=& - k_{Iy}, \non \\
k_{Ty} &=& \frac{1}{v_2} ~\sqrt{v_1^2 (k_{Ix}^2 + k_{Iy}^2) ~-~ v_2^2 
k_{Ix}^2}, \eea
assuming that the quantity inside the square root in the second equation is 
positive; otherwise $k_{Ty}$ will be imaginary and we will have total internal 
reflection on the left side of the line junction \cite{taka}. Thus 
${\vec k}_R$ and ${\vec k}_T$ are fixed in terms of ${\vec k}_I$. If we
define $\theta = \tan^{-1} (k_{Ix}/k_{Iy})$ and $\theta' = \tan^{-1} 
(k_{Tx}/k_{Ty})$, where $- \pi /2 \le \theta, ~\theta' \le \pi/2$, we have
\beq \frac{1}{v_1} ~\sin \theta ~=~ \frac{1}{v_2} ~\sin \theta'. \eeq
Finally, current conservation in the $\hat y$ direction at the line junction 
at $y=0$ implies that
\beq v_1 k_{Iy} ~( 1 ~-~ |r|^2) ~=~ v_2 k_{Ty} ~|t|^2. \eeq

The reflection and transmission amplitudes $r$ and $t$ can be obtained for
the general boundary condition in Eq.~(\ref{bc2}). The general expressions
for $r$ and $t$ are 
\bea t &=& \sqrt{\frac{v_1}{v_2}} ~\frac{\cos \theta ~e^{i(\theta' - 
\theta)/2}}{\cos \al ~\cos ( \frac{\theta + \theta'}{2} ) ~+~ i \sin \al~
\cos ( \frac{\theta - \theta'}{2})}, \non \\
r &=& e^{-i\theta} ~\frac{\sin \al ~\sin ( \frac{\theta + \theta'}{2} ) ~+~ i
\cos \al ~\sin ( \frac{\theta' - \theta}{2})}{\cos \al ~\cos ( \frac{\theta + 
\theta'}{2} ) ~+~ i \sin \al ~\cos ( \frac{\theta - \theta'}{2})}. 
\label{tr0} \eea
For equal velocities $v_1 = v_2$, we have $\theta' = \theta$ and 
Eq.~(\ref{tr0}) simplifies to
\bea t &=& \frac{\cos \theta}{\cos \al \cos \theta + i \sin \al}, \non \\
r &=& \frac{e^{-i\theta} \sin \al \sin \theta}{\cos \al \cos \theta + i 
\sin \al}. \label{tr1} \eea

Given the transmission amplitude $t$, we can compute the conductance across 
the line junction. This will be discussed in Sec. IV for a more general model 
in which the surfaces 1 and 2 are at an angle with respect to each other.

\subsection{Edge states}

It turns out that for certain ranges of values of $v_1/v_2$ and $\al$, there
are edge states which propagate along the line junction (i.e., along the 
$\hat x$ direction) with a momentum $k$ and whose wave functions decay 
exponentially as $y \to \pm \infty$. At the line junction, let us look for 
unnormalized wave functions of the form
\bea \psi_1 &=& \left( \begin{array}{c}
1 \\
\ga_1 \end{array} \right) ~e^{i(kx -Et) +\chi_1 y} ~~{\rm in ~region ~1},
\non \\
\psi_2 &=& \left( \begin{array}{c}
\ga_2 \\
\ga_3 \end{array} \right) ~e^{i(kx - Et) - \chi_2 y} ~~{\rm in ~region ~2}, 
\label{psie1} \eea
where $\ga_1, ~\ga_2, ~\ga_3$ can be complex numbers but the $\chi_i$ must 
be real and 
positive. The energy is given by $E = \pm v_1 \sqrt{k^2 - \chi_1^2} = \pm v_2
\sqrt{k^2 - \chi_2^2}$; we must have $\chi_1, ~\chi_2 \le |k|$. We now demand 
that the wave functions in Eq.~(\ref{psie1}) be eigenstates of Eq.~(\ref{ham1})
and that they satisfy Eq.~(\ref{bc2}); this gives us the values of the $\ga_i$ 
and $\chi_i$. Let us parametrize $\chi_1 /k = \sin \theta_1$ and $\chi_2 /k 
= \sin \theta_2$, where $-\pi/2 < \theta_1, ~\theta_2 < \pi/2$. We further 
assume that $v_1 \le v_2$, and define $\beta = \cos^{-1} (v_1/v_2)$, where 
$0 \le \beta < \pi/2$. 
We then find that edge states exist under the following conditions.
For $\beta/2 < \al < \pi/2$, the energy is negative and given by $E = - v_1 
\sqrt{k^2 - \chi_1^2}$, while for $- \pi/2 < \al < - \beta/2$, the energy is 
positive and given by $E = v_1 \sqrt{k^2 - \chi_1^2}$. In both cases, the
magnitude of the velocity, $|E/k|$, is given by
\beq v ~=~ \frac{v_1 v_2 |\sin (2\al)|}{\sqrt{v_1^2 + v_2^2 - 2 v_1 v_2 \cos
(2\al)}}. \label{vedge} \eeq
Further, 
\bea \tan \theta_1 &=& -~sgn (E/k) ~\frac{v_1 - v_2 \cos (2\al)}{v_2 \sin 
(2\al)}, \non \\
\tan \theta_2 &=& -~sgn (E/k) ~\frac{v_2 -v_1 \cos (2\al)}{v_1 \sin (2\al)}.
\eea
where $sgn$ denotes the signum function: $sgn (z) \equiv + 1$ if $z> 0$ and 
$-1$ if $z < 0$. Note that there are no edge states if $- \beta/2 \le \al \le 
\beta/2$.

We can use the above expressions to show that the velocity $v$ is smaller than
both $v_1$ and $v_2$. To state this differently, for a given value of the 
$\hat x$-momentum $k$, the energy of the edge states has a smaller magnitude 
than the energy of the plane waves in both regions 1 and 2. (At the two 
points $\al = \pm \beta/2$, we find that $v = v_1$).

\begin{figure} \begin{center} \epsfig{figure=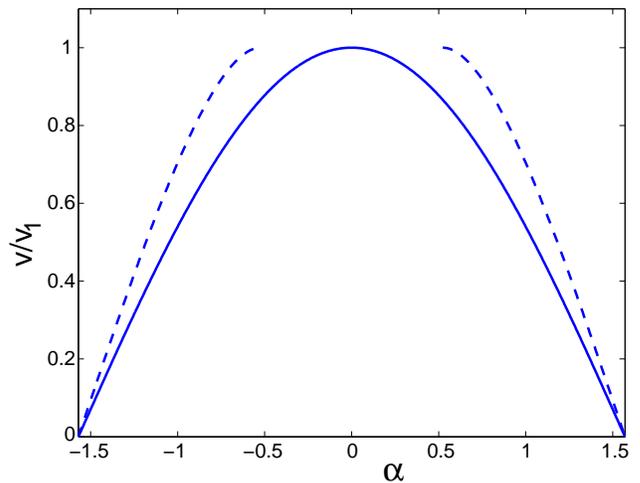,width=8.6cm} \end{center}
\caption{(Color online) Plot of $v/v_1$ versus $\al$ for $v_1/v_2 = 1$ 
(solid) and $0.5$ (dashed).} \label{fig2} \end{figure}

The various expressions above simplify if $v_1 = v_2$. Then there are edge 
states in the entire range $-\pi/2 < \al < \pi/2$, with $v= v_2 |\sin \al|$ 
and $\theta_1 = \theta_2 = - sgn (E/k) ~\al$. In Fig.~\ref{fig2}, we present a
plot of the velocity of the edge states as a function of $\al$, in the range
$-\pi/2 \le \al \le \pi/2$, for $v_1/v_2 = 1$ and $0.5$.

We emphasize that the sign of the energy $E$ of the edge states depends only 
on the sign of $\al$, and not on the sign of the momentum $k$. This is
consistent with the time reversal invariance of the system; under time
reversal, the momentum reverses sign but the energy remains the same.

After obtaining the values of the $\ga_i$ in Eq.~(\ref{psie1}), we can find 
the orientation of the spin of the edge states. For a spin-1/2 particle whose 
spin points along a unit vector ${\hat n} = (\sin a \cos b, \sin a \sin b, 
\cos a)$, we know that the normalized wave function is given by the transpose 
of $(\cos (a/2), \sin (a/2) e^{ib})$. Using this we find that the spin of the 
edge states points along the direction $- \sin \theta_1 ~{\hat z} \mp \cos 
\theta_1 ~{\hat y}$ in region 1 and along the direction $\sin \theta_2 ~{\hat 
z} \mp \cos \theta_2 ~{\hat y}$ in region 2, where the $\mp$ signs refer to 
$E > 0$ and $< 0$ respectively. Thus the spin of the edge states lies in
the $y-z$ plane and {\it not} in the $x-y$ plane which is the surface of
the TIs.

To conclude this section, we see that several properties of the edge states 
depend only on the dimensionless parameters $\al$ and $v_1/v_2$. This is true 
of the velocity $E/k$, the spin orientations in the two regions, and the 
product of $k$ and the exponential decay lengths $1/\chi_1$ and $1/\chi_2$.

\section{Junction between surfaces at an angle}
\label{sec3}

\begin{figure} \begin{center} \epsfig{figure=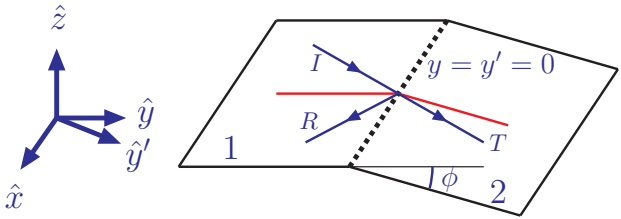,width=8.6cm} \end{center}
\caption{(Color online) Schematic picture of two regions, labeled 1 and 2, 
separated by a line junction at $y=y'=0$ where a thin barrier may be present; 
the unit vectors normal to the two regions are at an angle $\phi$ with respect
to each other. A wave incident ($I$) from region 1 and reflected ($R$) and 
transmitted ($T$) waves are shown.} \label{fig3} \end{figure}

We will now study what happens when there is a line junction between two 
surfaces which are boundaries of the same TI, but the surfaces are at an 
angle $\phi$ with respect to each other; this is shown in Fig.~\ref{fig3}. 
This is a realistic situation since any finite-sized TI must be bounded by 
surfaces meeting along lines. We will assume that the velocity $v$ is the 
same and is positive on both surfaces. Region 1 forms the $x-y$ plane, while 
region 2 forms the $x-y'$ plane where ${\hat y'} = \cos \phi ~{\hat y} - \sin 
\phi ~{\hat z}$. In regions 1 and 2, $y< 0$ and $y' > 0$ respectively; the 
line junction lies at $y=y'=0$, and we will again allow for a thin barrier 
to be present there. The Hamiltonians in the two regions are given by
\bea H_1 &=& - i v ~( \si^x \pa_y ~-~ \si^y \pa_x ), \non \\
H_2 &=& -i v ~( \si^x \pa_{y'} ~-~ \si^{y'} \pa_x ), \label{ham4} \eea
where $\si^{y'} = \cos \phi ~\si^y - \sin \phi ~\si^z$.
This model is of physical interest since any three-dimensional TI of finite 
size must necessarily have surfaces meeting along lines; our model describes 
the region around one such line. We will take $\phi$ to lie in
the range $-\pi < \phi < \pi$, although real systems may only allow a more
restricted range of $\phi$. The system described in Eq.~(\ref{ham4}) is 
invariant under the parity and time reversal transformations.

We may now ask what the general linear condition must be between 
$(\psi_1)_{y \to 0-}$ and $(\psi_2)_{y' \to 0+}$ so that the Hamiltonian of 
the complete system is Hermitian. Following an analysis similar to the one 
given in the Appendix for the model in Sec. II, we find that the general 
relation involves four parameters. But if we demand that the system be
invariant under time reversal, we find that only one parameter can appear,
namely, $(\psi_2 )_{y' \to 0+} = e^{-i\al \si^x} (\psi_1)_{y \to 0-}$.
This also ensures current conservation at the line junction, namely,
\beq (\psi_1^\dg \si^x \psi_1)_{y \to 0-} ~=~ (\psi_2^\dg \si^x \psi_2 )_{y'
\to 0+}. \label{curr3} \eeq

\subsection{Scattering}

For plane waves, the energy spectrum and wave functions in region 1 have the 
form given in Eq.~(\ref{psien1}), while in region 2, we obtain 
\bea \psi_+ &=& \frac{e^{-i\theta}}{\sqrt 2} ~\left( \begin{array}{c}
\cos \theta + i \sin \theta \cos \phi \\
1 - \sin \theta \sin \phi \end{array} \right)
e^{i(k_x x + k_y y' - E_+ t)}, \non \\
\psi_- &=& \frac{e^{-i\theta}}{\sqrt 2} ~\left( \begin{array}{c}
\cos \theta + i \sin \theta \cos \phi \\
- 1 - \sin \theta \sin \phi) \end{array} \right)
e^{i(k_x x + k_y y' - E_- t)}, \non \\
\label{psien2} \eea
where $E_\pm = \pm v \sqrt{k_x^2 + k_y^2}$ and $\theta = \tan^{-1} (k_x/k_y)$.
With the normalization given in Eq.~(\ref{psien2}), the currents in the $y'$
direction are given by $\psi_+^\dg \si^x \psi_+ = \cos \theta ~(1 - \sin 
\theta \sin \phi)$ and $\psi_-^\dg \si^x \psi_- = - \cos \theta ~(1 + \sin 
\theta \sin \phi)$.

We can then solve a scattering problem in which a wave is incident on the line
junction from region 1 with momentum ${\vec k}_I = (k_{Ix}, k_{Iy})$, and gets
reflected and transmitted with amplitudes $r$ and $t$ respectively. These 
amplitudes depend on the incident angle $\theta$, the bending angle $\phi$, 
and the parameter $\al$. The general expression for $t$ is
\bea t &=& \frac{2 \cos \theta}{A_1 \cos \al ~+~ i A_2 \sin 
\al}, \non \\
A_1 &=& \cos \theta + i \sin \theta \cos \phi + e^{-i\theta} (1 - \sin \theta 
\sin \phi), \non \\
A_2 &=& 1 - \sin \theta \sin\phi + e^{-i\theta} (\cos \theta + i \sin \theta 
\cos \phi). \label{tr2} \eea
We note that the transmission probability $|t|^2$ remains the same if we 
simultaneously change $\theta \to - \theta$, $\al \to - \al$ and $\phi \to 
- \phi$. As a result, the conductance (to be discussed in the next section) 
remains the 
same if we simultaneously change $\al \to - \al$ and $\phi \to - \phi$,
but generally changes if $\al \to - \al$ or $\phi \to - \phi$ separately.

\subsection{Conductance}

We will now compute the conductance across the line junction. Suppose that
the system is coupled at the far left ($y \to -\infty$) and far right 
($y' \to \infty$) to reservoirs with chemical potentials $\mu_L$ and $\mu_R$ 
respectively; we will assume that the reservoirs are at zero temperature. 
Electrons coming from one reservoir transmit across the line junction and go 
to the other reservoir with a probability given by $|t|^2$ given in 
Eq.~(\ref{tr2}). Assuming that the system has a large width in the $\hat x$ 
direction given by $W$, the net current going from left to right is given by
\beq I ~=~ e W ~\int \int~ \frac{dk_x dk_y}{(2\pi)^2}~ |t|^2 ~v \cos \theta ~
(1 - \sin \theta \sin \phi), \label{iv1} \eeq
where $e$ is the charge of an electron, and the factor of $v \cos \theta
(1 - \sin \theta \sin \phi)$ appears because we are interested in the $y'$ 
component of the transmitted current in region 2. For the integral 
in Eq.~(\ref{iv1}), the energy $E = \hbar v \sqrt{k_x^2 + k_y^2}$
goes from $\mu_R$ to $\mu_L$, assuming that $\mu_R < \mu_L$, and the
angle $\theta = \tan^{-1} (k_x /k_y)$ goes from $-\pi/2$ to $\pi/2$.
The voltage applied to a reservoir is related to its chemical potential as 
$\mu = q V$; in the zero bias limit, $\mu_L, \mu_R \to \mu$, we obtain the 
differential conductance
\beq G ~=~ \frac{dI}{dV} ~=~ \frac{e^2 W \mu}{v (2\pi \hbar)^2} ~
\int_{-\pi/2}^{\pi/2} ~d \theta~ |t|^2 ~\cos \theta ~(1 - \sin \theta \sin 
\phi). \label{iv2} \eeq
Using Eqs.~(\ref{tr2}) and 
(\ref{iv2}), we can calculate $G$ as a function of $\al$ and $\phi$. For 
$\al = \phi = 0$, we have $t=1$ and we obtain the conductance $G_0 = 2e^2 
W \mu/(v (2\pi \hbar)^2)$. 

\begin{figure} \begin{center} \epsfig{figure=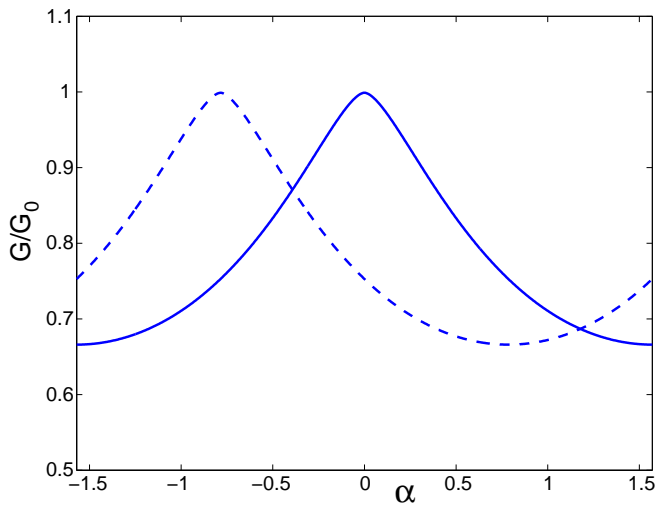,width=8.6cm} \end{center}
\caption{(Color online) Plot of $G/G_0$ versus $\al$ for $\phi = 0$ (solid) 
and $\pi/2$ (dashed).} \label{fig4} \end{figure}

\begin{figure} \begin{center} \epsfig{figure=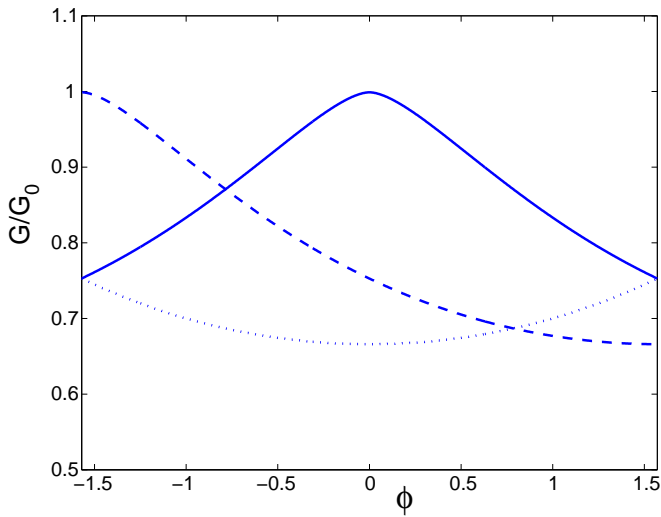,width=8.6cm} \end{center}
\caption{(Color online) Plot of $G/G_0$ versus $\phi$ for $\al = 0$ (solid), 
$\pi/4$ (dashed) and $\pi/2$ (dotted).} \label{fig5} \end{figure}

Fig.~\ref{fig4} shows a plot of the conductance (in units of $G_0$) as a
function of $\al$, in the range $-\pi/2 \le \al \le \pi/2$, for $\phi = 0$ 
(solid) and $\pi/2$ (dashed). Fig.~\ref{fig5} shows a plot of the conductance 
as a function of $\phi$, in the range $-\pi/2 \le \phi \le \pi/2$, for $\al 
= 0$, $\pi/4$ and $\pi/2$.
Interestingly, we find that the conductance is unity (in units of $G_0$)
on the line $\al = - \phi/2$, where $-\pi/2 \le \al \le \pi/2$.

\subsection{Edge states}

For any non-zero value of $\phi$, we find that there are edge states which 
propagate along the line junction with a momentum $k$ and whose wave 
functions decay exponentially as either $y \to - \infty$ or $y' \to \infty$. 
Let us set $\al =0$, so that we have the boundary condition $(\psi_2 )_{y' \to
0+} = (\psi_1)_{y \to 0-}$ at the line junction. We now look for unnormalized 
wave functions of the form
\bea \psi_1 &=& \left( \begin{array}{c}
1 \\
\ga \end{array} \right) ~e^{i(kx - Et) + \chi y} ~~{\rm in ~region ~1}, 
\non \\
\psi_2 &=& \left( \begin{array}{c}
1 \\
\ga \end{array} \right) ~e^{i(kx - Et) - \chi y'} ~~{\rm in ~region ~2}, 
\label{psie2} \eea
where $\ga$ can be complex but $\chi$ must be real and positive; the above 
wave functions satisfy the boundary condition given above. Demanding that 
these wave functions be eigenstates of Eq.~(\ref{ham4}), we find that there is
an edge state for any non-zero values of $k$ and $\phi$. Since $-\pi < \phi 
< \pi$ and $\phi \ne 0$, we can use the facts that $\cos (\phi/2)$ is always 
positive and that $\sin (\phi/2)$ and $\phi$ have the same sign. We then find 
that if $\phi > 0$, the edge state energy is negative and given by $E = - v 
\sqrt{k^2 - \chi^2}$, while if $\phi < 0$, the energy is positive and given 
by $E = v \sqrt{k^2 -\chi^2}$. In either case, we obtain the relation 
\beq \frac{\sqrt{k^2 - \chi^2}}{k-\chi} ~=~ \frac{\sqrt{k^2 - \chi^2} + k 
|\sin \phi|}{k \cos \phi + \chi}, \eeq
whose solution is given by
\beq \chi ~=~ | k ~\sin (\phi/2) |. \eeq
For instance, $\chi = |k \phi/2|$ for $\phi \to 0$ and $= |k|/\sqrt{2}$ for
$\phi = \pm \pi/2$. The quantity $\ga$ in Eq.~(\ref{psie2}) is given by 
\beq \ga ~=~ i ~\frac{\cos (\phi/2)}{sgn (k\phi) - \sin (\phi/2)}. \label{ga2} 
\eeq

We conclude that several properties of the edge states depend only on the 
bending angle $\phi$ and on the sign of the momentum $k$. The product of the
exponential decay length $1/\chi$ and $|k|$ is equal to $1/|\sin (\phi/2)|$. 
The magnitude of the velocity, $|E/k|$, is given by $v \cos (\phi/2)$. We may 
also consider the direction of the spin for the wave functions in 
Eq.~(\ref{psie2}) with $\ga$ given by Eq.~(\ref{ga2}). We find that the spin 
points along the direction ${\hat n} = \cos (\phi/2) ~{\hat y} - \sin (\phi/2)~
{\hat z}$ if $k\phi > 0$ and along $- {\hat n} = -\cos (\phi/2) ~{\hat y} + 
\sin (\phi/2) ~{\hat z}$ if $k\phi < 0$. It is interesting to note that 
$\hat n$ lies half-way between $\hat y$ and $\hat y' = \cos \phi ~{\hat y} - 
\sin \phi~ {\hat z}$.

\section{Junction between surface velocities with opposite signs}
\label{sec4}

Let us now study the case where there are two regions as in Sec. \ref{sec2},
but the velocities $v_1$ and $v_2$ in those two regions have opposite signs 
\cite{taka}. To be specific, let us assume that $v_1 > 0$ and $v_2 < 0$, and 
we send in a wave from region 1 with positive energy as before. Let us 
consider normal incidence, so that $k_{Ix} = k_{Rx} = k_{Tx} = 0$.
We then see that the operator $\si^x$ commutes with both $H_1$ and $H_2$.
For $k_{Iy} > 0$ and positive energy $E = v_1 k_{Iy}$, the incident wave 
function is an eigenstate of $\si^x$ with eigenvalue 1; hence the reflected 
and transmitted waves must also be eigenstates of $\si^x$ with eigenvalue 1.
However, this implies that the reflected wave must have $k_{Ry} = k_{Iy}$
and the transmitted wave must have $k_{Ty} = (v_1/v_2) k_{Iy}$; neither of 
this is possible since a reflected wave must have $k_{Ry} < 0$ and a 
transmitted wave must have $k_{Ty} > 0$. We therefore see that the
reflected and transmitted amplitudes must both vanish, leading to a 
difficulty in conserving the probability. 

One way out of this difficulty is to look for a linear relation
between $(\psi_1)_{y \to 0-}$ and $(\psi_2 )_{y \to 0+}$ which satisfies
current conservation as in Eq.~(\ref{curr1}). We find that the relation
\beq (\psi_2 )_{y \to 0+} ~=~ \sqrt{\frac{v_1}{|v_2|}} ~e^{-i\al \si^x} ~
\si^z ~(\psi_1)_{y \to 0-} \label{bc6} \eeq
works. However, there does not seem to be a physical way of justifying the 
factor of $\si^z$ on the right hand side of Eq.~(\ref{bc6}), although the 
factor of $e^{-i\al \si^x}$ can be justified in the same way as we did in the 
Appendix for the case of velocities with the same sign. In fact, the factor 
of $\si^z$ in Eq.~(\ref{bc6}) makes it non-invariant under the transformation 
$\psi_{1,2} \to \si^x \psi_{1,2}$ which was used above to show that the 
reflection and transmission amplitudes must vanish for normal incidence.
The factor of $\si^z$ would also break invariance under time reversal.

\begin{figure} \begin{center} \epsfig{figure=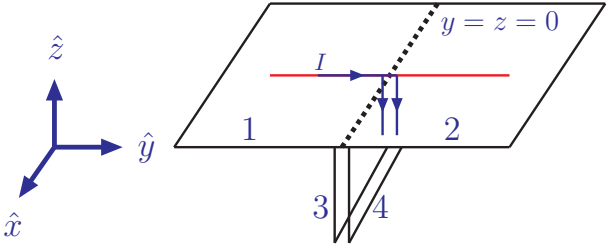,width=8.6cm} \end{center}
\caption{(Color online) Schematic picture of two regions of the $x-y$ plane, 
labeled 1 and 2, separated by a line junction at $y=z=0$, and two additional 
surfaces, labeled 3 and 4, which lie in the $x-z$ plane and are at the 
interface of the two TIs. A wave incident ($I$) from region 1 at $\theta=0$ 
and transmitted waves in regions 3 and 4 are shown.} \label{fig6} \end{figure}

An alternative and physically appealing way out of the difficulty in 
conserving probability was proposed in Ref. \onlinecite{taka}, by
realizing that in the case of opposite velocities, 
the waves must transmit {\it into} the interface regions 3 and 4 separating 
the two TIs, as depicted in Fig.~\ref{fig6}. Regions 3 and 1 will both
be taken to be surfaces of the first TI; we therefore assume that the 
velocity in both these regions is equal to $v_1$. Similarly, regions 4 and 
2 are both surfaces of the second TI and therefore have the same velocity 
$v_2$. The line junction is therefore a boundary of four surfaces, namely, 1, 
2, 3 and 4; the location of the line junction is given by $y=0$ for surfaces 
1 and 2, and $z=0$ for surfaces 3 and 4 (with $z$ becoming negative as one 
moves away from the line junction on these two surfaces). 
As we will discuss below, it will be convenient to assume that the line
junction has a small width separating the first and second TIs.
To see how a wave incident from region 1 can transmit into regions 3 and 4, 
we must first show that the interfaces 3 and 4 must necessarily have gapless 
states; otherwise energy conservation would not allow a very low-energy
incident wave in region 1 to transmit into the interfaces 3 and 4. Let us
therefore consider the Hamiltonian describing an electron moving on surfaces 
3 and 4. Since 3 and 4 both lie in the $x-z$ plane, but the unit vectors 
normal to those surfaces point in the $\hat y$ and $- \hat y$ directions 
respectively, the corresponding Hamiltonians are given by 
\bea H_3 &=& - i v_1 ~( \si^z \pa_x ~-~ \si^x \pa_z ), \non \\
H_4 &=& i v_2 ~( \si^z \pa_x ~-~ \si^x \pa_z ). \label{ham2} \eea

Next, we examine what boundary condition must be imposed at the line junction 
so that the complete Hamiltonian is Hermitian. Since the system consists of 
four Hamiltonians, $H_1, ~H_2, ~H_3$ and $H_4$, the general boundary condition
which ensures Hermiticity
will be much more involved than the one discussed in the Appendix for the case 
of two Hamiltonians. We will therefore consider the simplest possible boundary 
condition. Let us suppose that in the region of the line junction, the 
first TI (with surfaces 1 and 3) is separated from the second TI (with 
surfaces 2 and 4) by some distance; hence $\psi_3$ will be connected
to only $\psi_1$, and $\psi_4$ will be connected to only $\psi_2$. Let 
us further assume that there is no potential barrier between surfaces 1 and 3 
and between surfaces 2 and 4, so that the parameter $\al$ introduced in 
Sec. II is now equal to zero. We then obtain the simple conditions 
\beq \psi_3 ~=~ \psi_1 ~~{\rm and}~~ \psi_4 ~=~ \psi_2 \label{match} \eeq
at the line junction. (Note that this gives four equations since the $\psi_i$ 
are all two-component spinors). This condition automatically implies that the 
incoming and outgoing currents normal to the line junction are equal to each 
other, namely,
\beq v_1 \psi_1^\dg \si^x \psi_1 ~-~ v_2 \psi_2^\dg \si^x \psi_2 ~=~ v_1 
\psi_3^\dg \si^x \psi_3 ~-~ v_2 \psi_4^\dg \si^x \psi_4. \label{curr2} \eeq

Finally, we introduce tunneling between surfaces 3 and 4. Assuming rotational
invariance in the $x-z$ plane, the equations of motion take the form 
\cite{taka}
\bea i \pa_t \psi_3 &=& - i v_1 ~( \si^z \pa_x ~-~ \si^x \pa_z ) \psi_3 ~+~
g \psi_4, \non \\
i \pa_t \psi_4 &=& i v_2 ~( \si^z \pa_x ~-~ \si^x \pa_z ) \psi_4 ~+~
g \psi_3, \label{ham3} \eea
where $g$ denotes the tunneling amplitude; we will assume that $g$ is real and 
positive. In defining our model, we have made the conceptually simple 
assumption that the region of the line
junction is distinct from the tunneling region so that the tunneling does not
modify the boundary condition introduced in Eq.~(\ref{match}). (We note that 
the system described in Eq.~(\ref{ham3}) is invariant under the parity and 
time reversal transformations defined in Sec. \ref{sec2}).

To make further analytical progress, let us assume that $v_2 = - v_1$. To 
find the energy spectrum and wave functions, we take 
\beq \psi_3 ~=~ u_3 e^{i(k_x x + k_z z -Et)}, ~~{\rm and}~~ \psi_4 ~=~ u_4 
e^{i(k_x x + k_z z -Et)}, \eeq
where $E>0$ and $u_3$ and $u_4$ are two-component spinors. If $u_3$, $u_4$ are
eigenstates of the operator $\si^z k_x - \si^x k_z$ with eigenvalue $k = 
\sqrt{k_x^2 + k_z^2}$, the energy spectrum is given by $E = v_1 k \pm g$, 
while if $u_3$, $u_4$ are eigenstates of the same operator with eigenvalue 
$-k$, the energies are given by $E = - v_1 k \pm g$. We thus see that the 
energy vanishes, not at the origin $(k_x,k_z)=(0,0)$, but on the circle
$\sqrt{k_x^2 + k_z^2} = g/v_1$. This circle of points, along with $E=0$, 
is not invariant under Lorentz transformations which will be discussed
at the end of this section. Note that if $v_2$ had been positive and 
equal to $v_1$, the energy spectrum of Eq.~(\ref{ham3}) would have taken a
Lorentz invariant form as discussed below.

We now consider a wave incident on the line junction from region 1. Then
there are four possibilities; it can get reflected back to region 1 with
amplitude $r$, transmitted to region 2 with amplitude $t$, and transmitted to 
regions 3 and 4 with amplitudes $t'$ and $t''$ respectively. Since there are 
four amplitudes to be found and four matching conditions in Eq.~(\ref{match}),
the scattering problem can be solved. In general, all the four amplitudes
will be non-zero. However, let us now consider the case of normal incidence;
then we indeed find that $r=t=0$ as expected due to the conservation of
$\si^x$. To be explicit, the incident wave here is of the form
\beq \psi_1 ~=~ \frac{1}{\sqrt 2} ~\left( \begin{array}{c} 
1 \\
1 \end{array} \right) ~e^{i(ky - Et)}, \label{psi1} \eeq
where $E=v_1 k$. Conservation of energy equal to $E$, momentum along the 
$\hat x$ direction equal to zero, and $\si^x$ equal to 1 in all the regions
imply that the transmitted waves in regions 3 and 4 must be of the form
\bea \psi_3 &=& \frac{1}{2} ~[ t' \left( \begin{array}{c}
1 \\
1 \end{array} \right) e^{-ik'z} + t'' \left( \begin{array}{c}
1 \\
1 \end{array} \right) e^{-ik''z}] ~e^{-iEt}, \non \\
\psi_4 &=& \frac{1}{2} ~[ t' \left( \begin{array}{c}
1 \\
1 \end{array} \right) e^{-ik'z} - t'' \left( \begin{array}{c}
1 \\
1 \end{array} \right) e^{-ik''z}] ~e^{-iEt}, \non \\
& & \label{psi34} \eea
where $k' =(E/v_1) - g$ and $k'' =(E/v_1) + g$, and we have normalized the 
eigenstates corresponding to $k'$ and $k''$ to unity. The expressions in 
Eq.~(\ref{psi34}) are obtained by finding solutions of Eq.~(\ref{ham3}) 
with energy $E$ and positive group velocities $dE/dk'$ and $dE/dk''$ so that 
they propagate in the $- {\hat z}$ direction away from the line junction. Using
Eq.~(\ref{match}) along with Eqs.~(\ref{psi1}-\ref{psi34}) gives $t' = t''
= 1/\sqrt{2}$. Note that $|r|^2 + |t|^2 + |t'|^2 + |t''|^2 + = 1$ as desired.

The case of a general angle of incidence can be studied in a similar way.
We assume that the plane waves associated with the incident wave in region 1
with amplitude 1, the reflected wave in region 1 with amplitude $r$, the 
transmitted wave in region 2 with amplitude $t$, and transmitted waves in 
regions 3 and 4 with amplitudes $t'$ and $t''$ are given by $e^{i(k_x x + k_y
y - Et)}$, $e^{i(k_x x - k_y y - Et)}$, $e^{i(k_x x + k_y y - Et)}$,
$e^{i(k_x x - k' z - Et)}$ and $e^{i(k_x x - k'' z - Et)}$ respectively, where
$k_y > 0$, $E = v_1 \sqrt{k_x^2 + k_y^2}$, $k' = - \sqrt{(g - E/v_1)^2 - 
k_x^2}$, $k''= \sqrt{(g + E/v_1)^2 - k_x^2}$, and we have assumed that $0 < 
E/v_1 < g$ and $g \pm E/v_1 > |k_x|$. The above expressions and conditions 
ensure that the group velocity of the incident wave points toward the line 
junction but the group velocities of the other four waves point away from the
line junction as desired. We will not write down the corresponding
two-component wave functions here, but simply note that the boundary 
conditions in Eq.~(\ref{match}) will provide four equations which will 
determine the values of $r$, $t$, $t'$ and $t''$. In general, all these
four amplitudes will be non-zero. In the limit that the angle of incidence 
$\theta \to \pm \pi/2$, we find that $r \to -1$ and $t, ~t', ~t'' \to 0$.

We may return to the case of surface velocities with the same sign as
discussed in Sec. II, and ask how the results for scattering from the line 
junction would change in that system if one introduces additional surfaces 3 
and 4 coupled by a tunneling amplitude as in Fig.~\ref{fig6}. Considering 
Eq.~(\ref{ham3}) with $v_2 = v_1$, we find that a tunneling amplitude $g$ gives
rise to an energy gap, instead of producing gapless states as in the case of 
velocities with opposite signs. To be specific, the energy spectrum of 
Eq.~(\ref{ham3}) would be given by $E^2 = v_1^2 (k_x^2 + k_z^2) + g^2$, so 
that $|E|$ has a minimum value of $g$. (Note that this form remains invariant 
under Lorentz transformations based on the velocity $v_1$. For instance, 
$E^2 - v_1^2 (k_x^2 + k_z^2)$ remains invariant under 
the transformation $k_x \to (k_x - v E/v_1^2)/\sqrt{1 - v^2/v_1^2}$,
$E \to (E - v k_x)/\sqrt{1 - v^2/v_1^2}$ and $k_z \to k_z$, for any value of 
$v$ whose magnitude is smaller than $v_1$). As a result, an electron which is 
incident from region 1 with an energy less than $g$ will either reflect back 
to region 1 or transmit to region 2, and the surfaces 3 and 4 will not play a 
role in this scattering problem.

\section{Discussion}
\label{sec5}

To summarize, we have shown that a number of interesting phenomena can occur 
at a line junction separating two surfaces of topological insulators. Assuming
invariance under time reversal, a line junction can have an arbitrary 
parameter which makes the Hamiltonian Hermitian and is consistent with current
conservation; this parameter may be understood as arising from a potential 
barrier lying along the junction. The conductance through the line junction 
depends on this parameter, and there may also be edge states which propagate 
along the junction. For a junction which separates two surfaces which are at 
an angle with respect to each other, we have shown that the conductance 
depends on the bending angle, and there are also edge states which lie along 
the junction. Our analysis has been based on simple models in which the 
junction width $d$ has been taken to be zero; as a result, many properties of 
the edge states, such as their velocity and spin orientation, depend only on 
dimensionless parameters such as $\al$ and $\phi$ but not on the magnitude of 
the momentum $k$ along the edge. Realistic junctions would have a finite 
width; the effects of the width need to be studied.

The edge states which lie along the junction differ in two ways from the 
plane wave states which lie far from the junction. The velocity of the edge 
states is always less than the velocities of the plane waves, so that the 
energy of the edge states is smaller in magnitude from the energy of the 
plane waves for a given value of the momentum along the line junction. 
Secondly, the orientation of the spin of the edge states is generally
different from the orientation of the plane waves; in particular, the
spins of the edge states have a non-zero component in the direction 
perpendicular to the surfaces. These differences in the energy-momentum 
dispersion and the spin structure imply that it may be possible to observe 
various features of the edge states and to distinguish them from the plane 
waves (which reside far away from the line junction) by using spin-resolved 
angle-resolved photoemission spectroscopy and tunneling from a spin 
polarized STM tip placed close to the junction.

Finally, for the case in which the velocities on the two surfaces have 
opposite signs, we have explicitly demonstrated that a wave incident on the 
junction from either side must generally transmit into some additional 
surfaces which must emerge from the junction. 

Before ending, we would like to point out that our analysis has ignored the 
effects of the hexagonal warping of the dispersion which is known to occur in 
TIs \cite{fu}. It would be interesting to study the effects of warping on the 
conductance through the junction and on the edge states. We would also like 
to mention recent studies of scattering and bound states near ferromagnetic 
domain walls on the surface of a TI \cite{wickles}, transport across 
step junctions on a TI surface \cite{biswas,alos}, edge states induced by a 
magnetic field at junctions of two TI surfaces \cite{sitte}, bulk models 
for studying states at surfaces with arbitrary orientations \cite{zhang3},
and tachyon-like states at the interface of two TIs \cite{apalkov}.

\section*{Acknowledgments}

D.S. thanks the Department of Science and Technology, India for financial 
support under Grant No. SR/S2/JCB-44/2010, and Sourin Das, Arindam Ghosh, 
Sumathi Rao, Rahul Roy and Krishnendu Sengupta for useful discussions.

\appendix

\section{General boundary condition which ensures Hermiticity}

In this Appendix, we will discuss the general linear condition between
$(\chi_1)_{y \to 0-}$ and $(\chi_2)_{y \to 0+}$ (and similarly for $(
\chi'_1)_{y \to 0-}$ and $(\chi'_2)_{y \to 0+}$) which satisfies 
Eq.~(\ref{rel2}). Let us assume that
\bea (\chi_2)_{y \to 0+} &=& U ~(\chi_1)_{y \to 0-}, \non \\
(\chi'_2)_{y \to 0+} &=& U ~(\chi'_1)_{y \to 0-}, \label{rel3} \eea
where $U$ is an arbitrary $2 \times 2$ matrix. Then Eq.~(\ref{rel2}) will
be satisfied if 
\beq U^\dg \si^x U ~=~ \si^x. \label{rel4} \eeq
Using the fact that the identity matrix $I$ and the three Pauli matrices $\vec
\si$ form a basis for $2 \times 2$ matrices, we find that the general solution 
to Eq.~(\ref{rel4}) is given by
\beq U ~=~ \exp [-i \al \si^x - i \beta - \ga \si^y - \de \si^z], \label{rel5} 
\eeq
where $\al, ~\beta, ~\ga$ and $\de$ are real parameters; Eq.~(\ref{rel4}) 
follows from the fact that $U^\dg \si^x = \si^x U^{-1}$. (Eq.~(\ref{rel5}) may
be contrasted with the form of a unitary matrix $V$ which satisfies $V^\dg V 
= I$; such a matrix can be written in terms of four real parameters as $V = 
\exp [-i \al \si^x - i \beta - i \ga \si^y - i \de \si^z]$).

We will now present a model of a junction which provides a physical 
understanding of the four parameters appearing in Eq.~(\ref{rel5}). Consider 
a thin barrier of width $d$ extending from $y=0$ to $y=d$ where there is a 
constant term $V$ consisting of both potential and magnetic parts, 
\beq V (y) ~=~ A + B \si^x + C \si^z + D \si^y, \label{pot} \eeq
where $A, ~B, ~C$ and $D$ are real so that the Hamiltonian is Hermitian. We 
can think of $A$ as a potential, and $B, ~C$ and $D$ as being 
proportional to the three components of a magnetic field which has a Zeeman 
coupling to the spin of the electron. We will eventually be interested in the 
limit of a $\de$-function barrier, so that $A, ~B, ~C, ~D \to \infty$ and 
$d \to 0$ keeping $Ad, ~Bd, ~Cd$ and $Dd$ fixed.

The Hamiltonian in the region $0 < y < d$ is now given by
\beq H_d ~=~ -i v_2 ~( \si^x \pa_y ~-~ \si^y \pa_x ) ~+~ A + B \si^x + C 
\si^z + D \si^y. \label{ham5} \eeq
Let us look for a state with energy $E$ and momentum $k$ in the $\hat x$ 
direction; the corresponding wave function is of the form
\beq \psi (x,y,t) ~=~ f(y) ~e^{i(kx - Et)}. \eeq
In the region $0 < y < d$, $f(y)$ satisfies the equation
\bea && \left[ -i v_2 ~\si^x \pa_y ~-~ v_2 k \si^y ~+~ A + B \si^x + C
\si^z + D \si^y \right] f \non \\
&& =~ E f. \label{fy} \eea
Since we are interested in the limit $A, ~B, ~C, ~D \to \infty$, we can 
ignore in Eq.~(\ref{fy}) the terms of order $E$ and $k$ which are finite.
The solution for $f$ is then given by 
\beq f(y) ~=~ \exp [(y/v_2) (-iA \si^x -iB - C \si^y + D \si^z)]~ f(0), \eeq
which implies that
\beq f(d) ~=~ \exp [(d/v_2) (-iA \si^x -iB - C \si^y + D \si^z)]~ f(0). \eeq
We now take the limit $A, ~B, ~C, ~D \to \infty$ and $d \to 0$ keeping $Ad 
\equiv v_2 \al$, $Bd \equiv v_2 \beta$, $Cd \equiv v_2 \ga$ and $Dd \equiv 
- v_2 \de$ fixed. By superposing states with different values of $E$ and $k$, 
we then see that any wave function 
$\psi (x,y,t)$ must satisfy
\beq \psi(x,0+,t) ~=~ \exp [-i \al \si^x - i \beta - \ga \si^y - \de \si^z]~ 
\psi (x,0-,t). \label{rel6} \eeq
This is the general boundary condition proposed in Eqs.~(\ref{rel3}) and 
(\ref{rel5}).

Let us now impose invariance under time reversal. We then expect that only 
the potential $A$ can remain non-zero in Eq.~(\ref{ham5}) while the 
other three terms must be zero since they can be interpreted as arising from 
a magnetic field. To see this formally, we observe that under time reversal, 
we have to complex conjugate the Hamiltonians in Eq.~(\ref{ham1}) and wave 
functions, and transform the time $t \to - t$; in addition, we have to do a 
unitary transformation by $\si^y$ in order to maintain the form of the 
Hamiltonian. In other words, $i \pa_t \psi = H \psi$ goes to $i \pa_t \si^y 
\psi^* = \si^y H^* \si^y \psi^*$, where $\si^y H^* \si^y = H$. On applying
this transformation to Eq.~(\ref{ham5}), we see that $A$ can be non-zero
but $B,~C$ and $D$ must vanish if $V$ has to be time reversal invariant.
Hence the parameters $\beta, ~\ga$ and $\de$ must vanish in Eq.~(\ref{rel6}).
We are therefore left with the single parameter $\al$ given in Eq.~(\ref{bc2}).
 
If $\al \ne 0$, the wave functions satisfying Eq.~(\ref{bc2}) are 
discontinuous at $y=0$ even if $v_2 = v_1$. This is not surprising. We 
recall that for the Schr\"odinger equation which is second order in spatial 
derivatives, a $\de$-function potential barrier leads to a discontinuity in 
the first derivative of the wave function. For the Dirac equation which is 
first order in spatial derivative, a $\de$-function potential leads to a 
discontinuity in the wave function.

We would like to emphasize here that our aim has been to obtain 
the most general boundary condition consistent with certain Hamiltonians 
separated by a line junction. By doing so, we have included the effects of a 
thin barrier which could possibly be present at the junction. In a particular 
system of interest, the relevant parameters defining the boundary conditions
would have to be determined from either the microscopic parameters of the 
system or from experimental studies of either transport across the junction
or the edge states present there. For a given system (for instance, if there
is actually no junction present so that the junction width $d$ is exactly 
zero), it may turn out that all the four boundary parameters are equal to 
zero; however, this would only be a special case of our analysis.

To conclude, our analysis shows that in general four real parameters are 
required to completely specify the boundary condition; this generalizes the 
discussion in Refs. \onlinecite{taka,raoux,concha,peres} where no such 
parameters were considered. A similar situation is known to arise for the 
Schr\"odinger equation with a discontinuity at one point; the general 
boundary condition at that point which ensures Hermiticity and conserves 
the current also has four parameters as discussed in Refs. 
\onlinecite{carreau} and \onlinecite{harrison}.

\end{document}